\documentclass[twocolumn,amsmath,amssymb,floatfix,prl,showpacs]{revtex4}

\usepackage{graphicx}           

\def\nm{{\ {\rm nm}}}                       
\def\micron{{\ \mu{\rm m}}}                 

\def\Hz{{\ {\rm Hz}}}                       

\def\us{{\ \mu{\rm s}}}                     
\def\ms{{\ {\rm ms}}}                       
\def\second{{\ {\rm s}}}                    

\def\nK{{\ {\rm nK}}}                       

\def\uT{{\ \mu{\rm T}}}                     

\def\Rb87{^{87}\rm{Rb}}                 
\def\Li6{^{6}\rm{Li}}                   

\def\ex{{\mathbf e}_x}                            
\def\ey{{\mathbf e}_y}                            
\def\ez{{\mathbf e}_z}                            
\def\er{{\mathbf e}_r}                            
\DeclareMathAlphabet\mathbfcal{OMS}{cmsy}{b}{n}

\def\udc{{\uparrow,\downarrow}}
\def\duc{{\downarrow,\uparrow}}

\def\shorttimes{\!\times\!}                          
\def\shorteq{\!=\!}                          
\def\shortapprox{\!\approx\!}

\newcommand{\ket}[1]{\left|#1\right>}

\begin{document}

\title{Quenched binary Bose-Einstein condensates: spin domain formation and coarsening}

\author{S.~De}
\affiliation{Joint Quantum Institute, University of Maryland and 
National Institute of Standards and Technology, College
Park, Maryland, 20742, USA}

\author{D.~L.~Campbell}
\affiliation{Joint Quantum Institute, University of Maryland and 
National Institute of Standards and Technology, College
Park, Maryland, 20742, USA}

\author{R.~M.~Price}
\affiliation{Joint Quantum Institute, University of Maryland and 
National Institute of Standards and Technology, College
Park, Maryland, 20742, USA}

\author{A.~Putra}
\affiliation{Joint Quantum Institute, University of Maryland and 
National Institute of Standards and Technology, College
Park, Maryland, 20742, USA}

\author{B.~M.~Anderson}
\affiliation{Joint Quantum Institute, University of Maryland and 
National Institute of Standards and Technology, College
Park, Maryland, 20742, USA}

\author{I.~B.~Spielman}
\affiliation{Joint Quantum Institute, University of Maryland and 
National Institute of Standards and Technology, College
Park, Maryland, 20742, USA}
\date{\today}

\begin{abstract}
We explore the time evolution of two component Bose-Einstein condensates (BEC's), quasi-1D with respect to their spinor dynamics, following a quench from one component BEC's with a ${\rm U}(1)$ order parameter into two component condensates with a ${\rm U}(1)\shorttimes{\rm Z}_2$ order parameter.   In our case, these two spin components have a propensity to phase separate, \emph{i.e.}, they are immiscible. Remarkably, these spin degrees of freedom can equivalently be described as a single component attractive BEC. A spatially uniform mixture of these spins is dynamically unstable,  rapidly amplifing any quantum or pre-existing classical spin fluctuations. This coherent growth process drives the formation of numerous spin polarized domains, which are far from the system's ground state. At much longer times these domains grow in size, coarsening, as the system approaches equilibrium.  The experimentally observed time evolution is consistent with our stochastic-projected Gross-Pitaevskii calculation.
\end{abstract}


\maketitle

\begin{figure}[t!]
\center
\includegraphics[width=3.3 in, angle = 0]{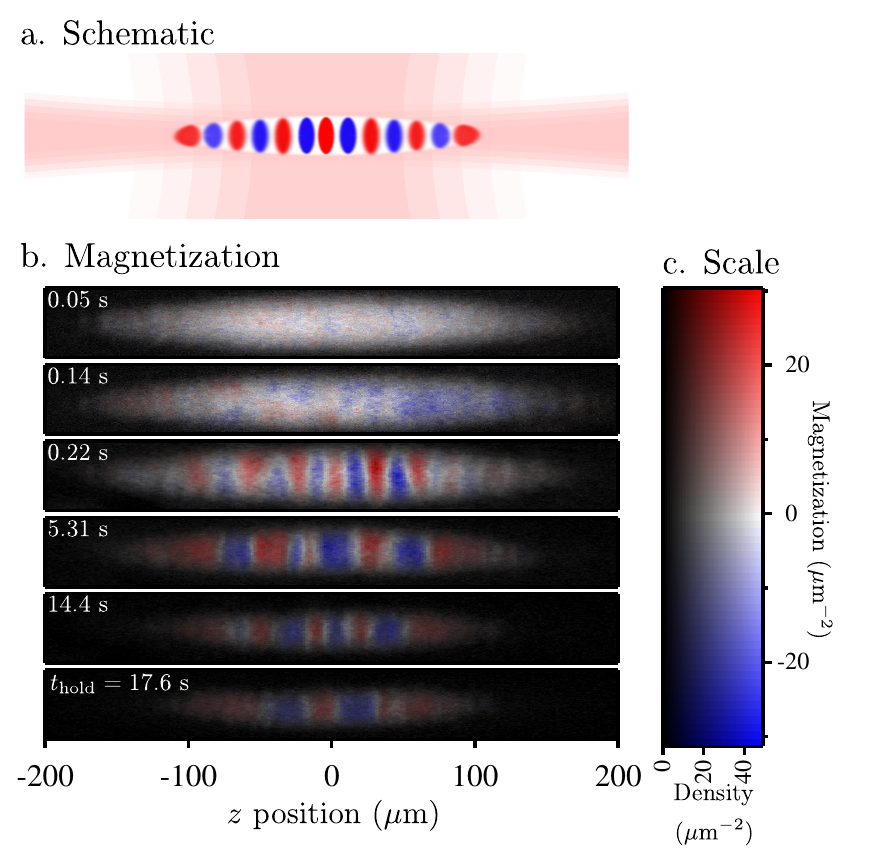}
\caption{
Magnetization $M_z({\bf r})$.  (a) Schematic, illustrating a spinor BEC with domains in an anisotropic crossed-dipole trap. (b) Images showing the progression from a uniformly magnetized condensate (short times) in which domains appear (intermediate times), and then grow spatially (long times); during this process the condensate slowly decays away.  (c) Color scale indicating the degree of magnetization (colors from blue to red), and the density (intensity from black to colored).
} \label{Fig:DomainDataPlot}
\end{figure}

Ultracold atomic gases are unique systems for studying phase
transitions where the full range from adiabatic to diabatic can be
easily accessed in the laboratory.  A prime example of this is the
transition from superfluid (SF) to Mott-insulator (MI) in an
optical lattice: when the transition is crossed slowly, a nearly
$T\shorteq0$ SF transforms into a nearly $T\shorteq0$ MI~\cite{Greiner2002};
however, when the system is quenched by rapidly entering the MI
regime, it exhibits rapid dynamics before dephasing into a highly
excited, high temperature final
state~\cite{Greiner2002a,Sebby-Strabley2007,Will2010}. Here we
study a similar quantum quench in a two component spinor BEC,
where the spin degree of freedom is initialized in a highly excited
state.  We follow the resulting dynamics during which spin
domains rapidly form, and subsequently slowly relax towards
equilibrium as the domain size increases and the domain number
decreases (see Fig~\ref{Fig:DomainDataPlot}).

The establishment of out of equilibrium domains formed by quenching through a phase transition is ubiquitous in physical systems ranging from grain formation in minerals~\cite{Putnis1992}, domain nucleation in magnetic systems, to Kibble-Zurek phenomena such as structure growth in the early universe~\cite{Kibble1976}, and spontaneous vortex formation in quenched BEC's~\cite{Weiler2008a}.  For an initially zero-temperature system, a quench can result from rapidly traversing a second order quantum phase transition that is associated with a change in the system's symmetry.  In the case of a quench from SF to MI, the SF's ${\rm U}(1)$ order parameter is absent in the MI phase (with its trivial ${\rm Z}_1$ order parameter).  In contrast, the order parameter transforms from ${\rm Z}_1$ to ${\rm Z}_2$ for a quenched transverse-field Ising ferromagnet~\cite{Calabrese2012}.  In our experiment, we prepare a transversely magnetized two component spinor BEC described by a ${\rm U}(1)$ order parameter, and observe the formation and spatial expansion (coarsening) of domains following a quench into a phase with a ${\rm U}(1)\shorttimes {\rm Z}_2$ order parameter~\cite{Barnett2006,Kawaguchi2011}, unexplored by previous studies with binary condensates (miscible~\cite{Weld2009,Hoefer2011} or immiscible~\cite{Hall1998,Mertes2007}).  As compared with three component systems~\cite{Stenger1998,Zhang2005a,Sadler2006a,Vengalattore2010,Guzman2011,Bookjans2011}, the relative simplicity present here allows us to identify an intriguing analogy between our spin system and a single-component {\it attractive} BEC as it collapses~\cite{Strecker2002,Pollack2009,Donley2001,Dutton2005}.

We explore the time-evolving magnetization of two-component $\Rb87$ Bose-Einstein condensates (BEC's) in the $5{\rm S}_{1/2}$ electronic ground state.  Our BEC's are well described in terms of a spinor wave-function $\Psi({\bf r}) \shorteq \left\{\psi_\uparrow({\bf
r}),\psi_\downarrow({\bf r})\right\}$, where the
$\ket{\uparrow,\downarrow}$ pseudo-spins label
the $\ket{f\shorteq1,m_F\shorteq\pm1}$ atomic spin states.  It's dynamics are given by the spinor Gross-Pitaevskii equation
(sGPE)
\begin{align}
i\hbar \partial_t \psi_{\udc}({\bf r}) =& \bigg[-\frac{\hbar^2
\nabla^2}{2m} + V({\bf r}) +\left(c_0-c_2\right) n({\bf r}) \label{eq:sGPE}\\
& + 2 c_2 N \left|\psi_{\udc}({\bf r})\right|^2
\bigg]\psi_{\udc}({\bf r}) + \frac{\Omega_\perp}{2}\psi_{\duc}({\bf r}),\nonumber
\end{align}
a continuum analog to the transverse field Ising model.  $n({\bf r}) \shorteq N\left[\left|\psi_\uparrow({\bf r})\right|^2 + \left|\psi_\downarrow({\bf r})\right|^2\right]$ is the total density; $m$ is the atomic mass; $V(\bf{ r})$ is a
spin-independent external potential; $\Omega_\perp$ describes the Zeeman shift of a ``transverse'' magnetic field; and $c_{0,2}$ are the spin-independent and spin-dependent interaction coefficients~\cite{Ho1998,Ohmi1998}.  This Hamiltonian has a ${\rm Z}_2$ symmetry describing a reversal of $\ket{\uparrow}$ and $\ket{\downarrow}$, which is absent in most binary mixtures~\cite{Weld2009,Hoefer2011,Hall1998,Mertes2007}.  In $\Rb87$'s $f\shorteq1$
manifold, $c_0 \shorteq (100.86)\shorttimes 4\pi\hbar^2 a_{\rm B}/m$ vastly exceeds $c_2 \shortapprox - 4.7\shorttimes10^{-3} c_0$, where $a_{\rm B}$ is the Bohr radius~\cite{vanKempen2002}.  For a static density profile and when $\Omega_\perp=0$, each spin component in Eq.~\eqref{eq:sGPE} is separately described by an attractive single-component GPE.

We produce $N\shorteq 7.0(5) \shorttimes10^5$ atom $\Rb87$ BEC's~\footnote{All uncertainties herein reflect the uncorrelated combination of single-sigma statistical and systematic uncertainties.} in the $\ket{f\shorteq1,m_F\shorteq0}$ hyperfine state, originating from cold $\ket{f\shorteq1,m_F\shorteq-1}$ thermal clouds formed in a hybrid magnetic/optical trap~\cite{Lin2009}.  These BEC's are subject to a uniform magnetic field with magnitude $B_0 \shorteq 107.0(2)\uT$ and are confined in the extremely anisotropic crossed optical dipole trap depicted in Fig.~\ref{Fig:DomainDataPlot}a.  Our dipole trap is formed from a pair of axially symmetric $1064\nm$ laser beams intersecting at right angles with $1/e^2$ radii $\shortapprox67\micron$ and $\shortapprox300\micron$. The radial ($\er$, i.e., in the $\ex - \ey$ plane) and axial ($\ez$) trap frequencies are $\omega_{r}/2\pi \shorteq 135(3)\Hz$ and $\omega_z/2\pi \shorteq 3.1(2)\Hz$ respectively.   Our $T\shorteq90 (8)\nK$ condensates have radial and axial Thomas-Fermi radii of $R_{r}\shorteq3.9(1)\micron$ and $R_{z}\shorteq170(7)\micron$. The BECs' $170\micron$ axial radius is not small compared to dipole laser's $300\micron$ waist along the axial direction; as a result, we expect small deviations from the conventional inverted parabola density profile.

Because the typical $c_0 n({\bf r})$ spin-independent energy vastly exceeds the $c_2 n({\bf r})$ spin-dependent energy scale, we make the conventional Thomas-Fermi approximation for the overall density distribution $n({\bf r})$ characterized by a chemical potential $\mu$, and a minimum healing length $\xi\shorteq\hbar/\sqrt{2 m \mu}$.  This gives $n({\bf r})\shorteq\left[\mu - V({\bf r})\right]/\left[c_0+c_2 M_z^2({\bf r})\right]$, which depends very weakly on the $z$ component of local magnetization vector, ${\bf M}({\bf r}) \shorteq \left\{M_x({\bf r}),M_y({\bf r}),M_z({\bf r})\right\} \shorteq \left\{2 {\rm Re}[\psi_\uparrow^*({\bf r})\psi_\downarrow({\bf r})], 2 {\rm Im}[\psi_\uparrow^*({\bf r})\psi_\downarrow({\bf r})], \left|\psi_\uparrow({\bf r})\right|^2 - \left|\psi_\downarrow({\bf r})\right|^2\right\}$.

The spin degrees of freedom vary almost exclusively with axial position~\cite{Bookjans2011} because our extremely anisotropic condensate's $\shortapprox\!3.9\micron$ radial extent is comparable to the minimum spin healing length $\xi_s \shorteq \xi|c_0/c_2|^{1/2} \shorteq 3.20(4)\micron$.  Theoretically, we may describe the spin degree of freedom as 1D spinor~\cite{Dutton2005} with components $\chi_{\udc}(z) \shorteq \left|\chi_{\udc}(z)\right|e^{i \phi_\udc(z)}$; retaining terms through first order in $c_2/c_0$, we obtain an effective 1D sGPE 
\begin{align}
i\hbar \partial_t \chi_{\udc} &=\! \left[ -\frac{\hbar^2 \partial_z^2}{2m}\!-\! g_{1{\rm D}}(z) + 2 g_{1{\rm D}}(z) \left|\chi_{\udc}\right|^2 \right] \chi_{\udc}\label{eq:1DsGPE}.
\end{align}
The 1D interaction strength $g_{\rm 1D}(z)\propto c_2$ is related to a 1D healing length $\xi_{1{\rm D}} \shortapprox \sqrt{3/2} \xi_{s}$.  These two 1D sGPE's are coupled by the local constraints $|\chi_{\uparrow}(z)|^2 + |\chi_{\downarrow}(z)|^2 \shorteq 1$ and $\phi_{\uparrow}(z) + \phi_{\downarrow}(z) \shorteq 0$ (i.e., no mass currents in our experiment). To make the analogy explicit, we dropped terms quadratic in $|\chi_{\udc}|^2$ resulting from integrating out the transverse dimensions. These repulsive terms do not affect the dynamics at short times after the quench, but must be included at long times. 

Our spinor experiment is initiated by a $34\us$ rf-pulse that puts each atom into a equal-amplitude superposition of the $\ket{\uparrow,\downarrow}\shorteq|m_F\shorteq\pm1\rangle$ spin states, the ground state when $\Omega_\perp$ is large; the system then evolves according to Eq.~\eqref{eq:sGPE} with $\Omega_\perp=0$.  This procedure is equivalent to rapidly quenching $\Omega_\perp$ to zero:  the ground state goes from breaking a ${\rm U}(1)$ symmetry to breaking a {\em different} ${\rm U}(1)$ along with a ${\rm Z}_2$ symmetry. While a conventional BEC breaks just a single ${\rm U}(1)$ symmetry associated with a wave function's overall phase (generated by the identity), our spinor Hamiltonian adds a ${\rm U}(1)$ symmetry associated with the relative phase of the spin (generated by the Pauli matrix $\check\sigma_z$), as well as a discrete ${\rm Z}_2$ symmetry. Post quench, the formation of spin domains corresponds to breaking the ${\rm Z}_2$ symmetry, while within a specific domain, a new ${\rm U}(1)$ symmetry is broken. This is generated by a combination of the overall and relative phases: each spin domain has a broken generator $(\check{1}\pm\check\sigma_z)/2$, leaving behind a ``sneaky'' unbroken ${\rm U}(1)$ symmetry generated by $(\check{1}\mp\check\sigma_z)/2$.

The quenched binary mixture is held for a variable duration $t_{\rm hold}$, up to $20 \second$, while spin structure forms and evolves.  Spin mixing collisions are suppressed because the relatively large $82\Hz$ quadratic Zeeman shift greatly exceeds the $c_2 n({\bf r})\shortapprox 6\Hz$ spin dependent energy~\cite{Stenger1998}.  As a result, we observe no population in $m_F\shorteq0$ for the entire duration of our experiment.  After $t_{\rm{hold}}$, we remove the confining potential and allow the atomic ensemble to expand (largely transversely) for $19.3 \ms$, during which time we Stern-Gerlach~\cite{Gerlach1922} separate the spin components. We detect the resulting density distribution by absorption imaging, and reconstruct both $M_x(x,z)$ and $M_z(x,z)$, projected onto the $\ez\!-\!\ex$ imaging plane. A brief rf pulse just before TOF can partially re-populate $\ket{m_F\shorteq0}$; following TOF expansion and Stern-Gerlach seperation, the distribution of all three spin states contains sufficient information to obtain $M_x$ and $M_z$ simultaneously. We depict representative reconstructions of $M_z(x,y)$ at six hold times in Fig.~\ref{Fig:DomainDataPlot}b.

\begin{figure}[b!]
\begin{center}
\includegraphics[width=3.3in]{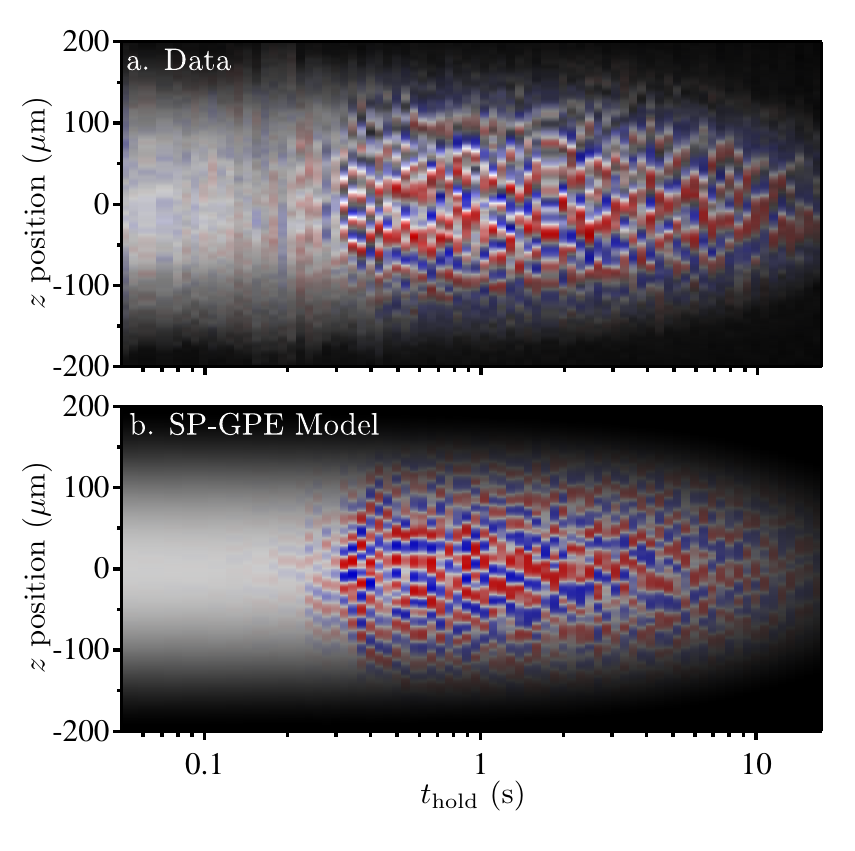}
\end{center}
\caption{Time evolution of magnetization $M_z(z)$. (a) Experimental data and (b) finite temperature simulation using the SP-GPE method.  In both simulation and experiment,  the spatial structure of $M_z(z)$ coarsens after an initial growth period as domains coalesce.  
} \label{fig:onset}
\end{figure}

The initially ($t_{\rm hold}\shorteq0$) uniform ${\boldsymbol \chi}(z) \shorteq \left(\ket{\uparrow} + \ket{\downarrow}\right)/\sqrt{2}$ spin superposition is dynamically unstable (as indicated in Fig.~\ref{Fig:DomainDataPlot}b's snapshots).  At this unstable point, small spin-wave excitations have an $\left(\hbar\omega/\mu_{\rm 1D}\right)^2 \shorteq  \left(k\xi_{\rm 1D}\right)^2 \left[ \left(k\xi_{\rm 1D}\right)^2 - 2 \right]$ energy spectrum~\cite{Stamper-Kurn2012}, where $\mu_{\rm 1D}\shorteq\hbar^2 / 2 m \xi^2_{\rm 1D}$ is a typical 1D spin interaction energy.  When $\hbar\omega$ is imaginary -- for $k\xi_{1{\rm D}}\in\left(0,\sqrt{2}\right)$ -- the associated modes grow exponentially with peak gain at $k\shorteq1/\xi_{1{\rm D}}$, amplifying any existing spin fluctuations, classical or quantum.  Figure~\ref{fig:onset} depicts the magnetization $M_z(z)$, showing the initially unmagnetized condensate develop visible structure after about $200\ms$.  The experimental data plotted in Fig.~\ref{fig:onset}a is in essentially perfect agreement with a stochastic-projective GPE (SP-GPE) simulation~\cite{Blakie2008}, with parameters nearly matched to our experiment, Fig.~\ref{fig:onset}b.  The SP-GPE's stochastic noise term was chosen to match the experimentally observed temperature, and was not tuned to match the onset-time for domain formation.

\begin{figure}[tb]
\center
\includegraphics[width=3.3 in]{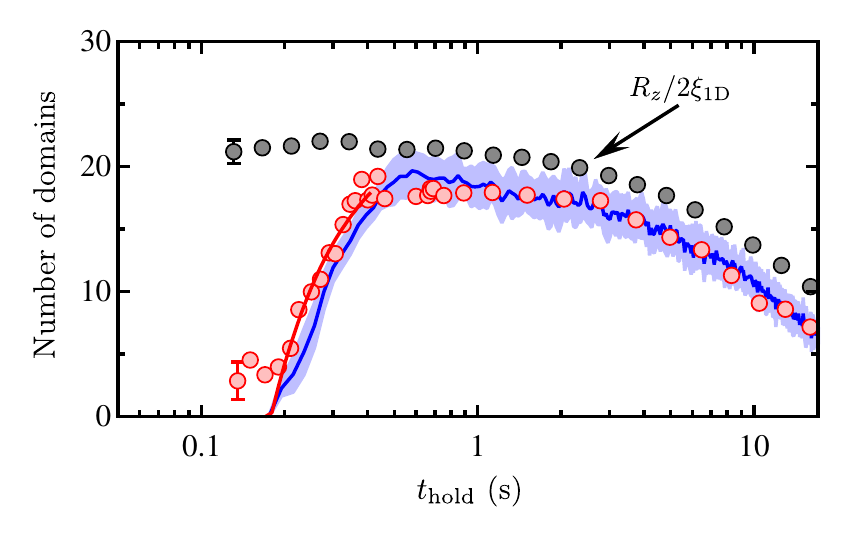}
\caption{
Number of domains as a function of $t_{\rm hold}$.  The red symbols depict the experimentally observed number of domains (typical uncertainty plotted on the leftmost point) and the blue curve plots the results of our SP-GPE simulation (uncertainties denoted by the blue band).  In both cases, the uncertainties reflect the standard deviation over many realizations.  In addition, the red curve fits the data to a model assuming exponential growth along with a non-zero observation threshold, in the LDA. The grey symbols correspond to the ratio $R_z/2\xi_{\rm 1D}$: an estimate of domain number, assuming the system with length $2R_z$ is partitioned into domains of local size $\pi\xi_{\rm 1D}(z)$ (the size at which domains initially form); the weighted average of this over our system is about $4\xi_{\rm 1D}$.
} \label{Fig:DomainNumber} 
\end{figure}

While the amplitude of these spin waves grow with an exponential time constant $\tau(k) \shorteq 1/{\rm Im}(\omega(k,z))$ [minimum at $\tau(z) \shorteq 2 m \xi^2_{\rm 1D}(z) / \hbar\shortapprox 42\ms$], Fig.~\ref{fig:onset} shows that no structure is visible until $t_{\rm hold}\shortapprox200\ms$.  Our simulations confirm that structure begins to grow immediately, and only technical noise prevents us from detecting the growing spin modulation at shorter times.  Figure~\ref{fig:onset} also shows that spin structure forms more slowly in the lower density periphery of the system where $\xi_{\rm 1D}$ and $\tau$ are larger.  To quantify this effect, Fig.~\ref{Fig:DomainNumber} plots the number of spin-regions visible above the noise, along with the results of our SP-GPE simulations, and a local density approximation (LDA, accounting for our systems inhomogeneous density profile) prediction for the expected pattern of domain growth.  This number increases for short times because spin-regions become visible in the system's center before its edges, and does not initially reflect a change of their spatial size.

The spin modulations continue to grow in amplitude until, at $t_{\rm hold}\shortapprox 300\ms$, they form fully spin polarized domains of $\ket{\uparrow}$, and $\ket{\downarrow}$, with a spacing set by the dynamic growth process, not by the system's equilibrium thermodynamics.  After this period of rapid growth, the polarized spin domains evolve slowly, equilibrating, for the remaining $20\second$ duration of our experiment.

Our BEC has a $\tau \shorteq 10(1)\second$ lifetime, implying that the domain pattern {\it must} evolve in time as the BEC slowly contracts.  The simplest model -- in which the domain pattern contracts together with the dwindling BEC (where each domain simply contracts) -- is obviated by Fig.~\ref{Fig:DomainNumber}, that shows the number of domains decreasing after $t_{\rm hold} \shortapprox 1\second$.  Indeed, once a domain becomes smaller than $\shortapprox 2\xi_{\rm 1D}(z)$, it can no longer reach full spin-polarization in its center, and it ceases to be a barrier for the hydrodynamic flow of the other spin state. As a result, small domains de-pin and can move freely until they coalesce with another domain of the same spin.

\begin{figure}[bt]
\begin{center}
\includegraphics[width=3.3in]{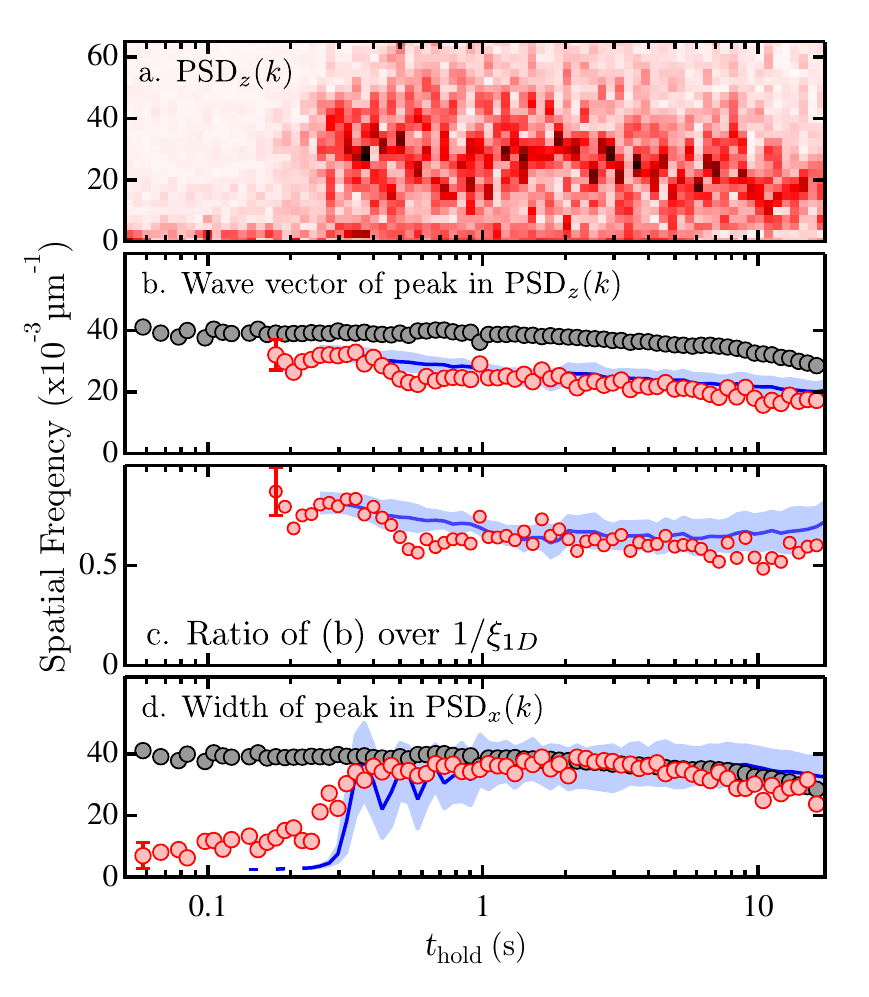}
\end{center}
\caption{Power spectral density. (a) ${\rm PSD}_{z}(k)$ as a function of $t_{\rm hold}$ showing the formation of a peak at finite wave-vector $k$, followed by the gradual movement of that peak to smaller $k$ as the spin domains expand. Each vertical slice represents a single experimental realization, i.e., no averaging.  The color scale depicts increasing spectral power with darker color.  (b) Wave-vector of ${\rm PSD}_{z}(k)$'s peak.    (c)  Ratio of ${\rm PSD}_{z}(k)$ peak wave-vector over $1/\xi_{\rm 1D}$.    (d) Width of ${\rm PSD}_{x}(k)$, which always peaked around zero.  In (b), (c) and (d), the red symbols depict the experimentally observed peak location (typical uncertainty plotted on the leftmost point) and the blue curve plots the results of our SP-GPE simulation (uncertainties denoted by the blue band).  In these three cases, the uncertainties reflect the standard deviation over eleven realizations, i.e., (b), (c) and (d) are averaged data.  The grey symbols mark $1/\xi_{\rm 1D}$, the homogenous-system wave-vector of maximum gain (the uncertainties are comparable to the symbol size).  The oscillations for $t_{\rm hold}<1\second$ in the simulation (blue curve, panel d) result from a damped breathing mode predominately along $\ez$.
} \label{fig:PSD}
\end{figure}

While Figs.~\ref{fig:onset} and \ref{Fig:DomainNumber} qualitatively suggest that the domains gradually expand as $t_{\rm hold}$ increases from $300\ms$ to $20\second$, it is difficult to obtain a quantitative measure of domain size from data in this form.  Indeed, the data show that while measurements at neighboring times have similar domain sizes, the exact domain pattern has a significant element of randomness -- primarily in the form of phase shifts -- likely resulting from subtle differences in the initial conditions, as amplified by the subsequent exponential gain process.  To mitigate these effects, we turn to the power spectral density ${\rm PSD}_{x,z}(k)\shorteq\left|\int M_{x,z}(z) \exp(i k z) {\rm d}z\right|^2$ obtained from these data.  With the PSD, we can compare different realizations even in the presence of spatial phase shifts of the domain structure.

Figure~\ref{fig:PSD}a shows ${\rm PSD}_z(k)$ derived from $M_z(z)$ shown in Fig.~\ref{fig:onset}.  For short times ($t_{\rm hold} \lesssim 300 \ms$), a narrow peak associated with the growing spin modulations develops.  Once the spin domains reach unity polarization, the magnetization's magnitude saturates and the boundaries between domains -- domain walls -- sharpen, broadening ${\rm PSD}_z(k)$ starting at $t_{\rm hold}\shortapprox300\ms$.  At longer times, the broad peak drifts to smaller wave-vector, indicating an increasing typical domain size.  Figure~\ref{fig:PSD}b compares this peak location for both experiment and theory (red and blue symbols respectively, showing nearly identical behavior) against $1/\xi_{\rm 1D}$.  Our simple model predicts maximum gain at this wave-vector, however, the peak in ${\rm PSD}_z(k)$ for both the experiment and the SP-GPE is at slightly smaller $k$.  This results from the inhomogenous density profile of the spinor BEC: simulations of uniform systems do show peak gain at  $1/\xi_{\rm 1D}$.

Because the $\;\shortapprox 2\xi_{\rm 1D}(z)$ minimum domain size increases as the condensate depletes away, it is plausible that the increase in domain-size results exclusively from an increasing cutoff in the minimum domain size. That this model is not fully consistent with the data can be seen by comparing $\xi_{\rm 1D}^{-1}$ computed in the BEC's center (Fig.~\ref{fig:PSD}, grey symbols) to the wave-vector of the peak in ${\rm PSD}_{z}(k)$; Fig.~\ref{fig:PSD} plots experimental data with red symbols and SP-GPE simulation with the blue curve.  They follow somewhat different time-dependences, with the domains growing rapidly at short times (from $400 \ms$ to $800\ms$) before $\xi_{\rm 1D}^{-1}$ appreciably changes and growing relatively little at long times ($t_{\rm hold} > 10\sec$) as $\xi_{\rm 1D}^{-1}$ falls more rapidly (Fig.~\ref{fig:PSD}c, present in the data and more dramatically in the simulation), indicating that $\xi_{\rm 1D}^{-1}$ is not simply related to the typical domain size.  In simulations for temperatures above the experimental temperature, rapid domain growth between $400 \ms$ and $800\ms$ often appeared due to solitons traversing the system, however, we have no direct experimental evidence to support this model.

Thus, a genuine coarsening of the domains, resulting from dynamical exchange of particles between stable domains partially contributes to the domain growth process.  On the basis of our data, we cannot identify the origin of this coarsening, but as with previous spinor BEC experiments~\cite{McGuirk2003}, our SP-GPE simulations suggest that spin transport through the uncondensed  fraction, not tunneling, is the leading mechanism.

Unlike ${\rm PSD}_z(k)$, ${\rm PSD}_x(k)$ is peaked about zero; this is because $M_x(z)$ is only appreciable in the domain walls where the gas is not fully polarized: it is a series of narrow peaks.  By showing that the width of the peak in ${\rm PSD}_x(k)$ tracks the inverse spin-healing length, Fig.~\ref{fig:PSD}d demonstrates that the domain walls are sized according to $\xi_{\rm 1D}$ (grey symbols).

In experiment, we saw a small, but repeatable large-scale structure in $M_z(z)$ due to a very small residual gradient $\nabla_z B \shorteq \gamma z$, where $\gamma \shorteq 0.092(4)\ {\rm G}/{\rm cm}^2$.  Although small, we could further mitigate the effects of this contribution with a spin echo $\pi$-pulse, flipping between $\ket{\downarrow}$ and $\ket{\uparrow}$ midway between the quench and the beginning of TOF.  This removed the large-scale spin structure from the inhomogeneous magnetic fields, but left the spin dynamics -- which were associated with much shorter length scales -- otherwise unaffected.  Furthermore, well above our ambient gradients, a counterflow instability~\cite{Hoefer2011} seeds spin-structure growth.

For $c_2\!<\!0$, as in $\Rb87$, Eqs.~\eqref{eq:sGPE} and~\eqref{eq:1DsGPE} describe our system's spin degree of freedom as a single component attractive BEC (the overall density follows the conventional Thomas-Fermi profile).  The process of domain formation is a spinor analog to the ``chain of pearls'' pattern that forms in 1D BEC's quenched from repulsive to attractive interactions~\cite{Strecker2002,Pollack2009}.  In that case, the growth of structure results from a modulational instability with peak gain at $k\shorteq1/\xi$ set by the conventional healing length.  Attractive Bose systems are intrinsically unstable against collapse~\cite{Donley2001}, however for spinors, any eventual collapse is stymied by an effective hard core interaction resulting from the bounded individual spin wavefunctions, and higher order interaction terms omitted from Eq.~\eqref{eq:1DsGPE}.

Thus, we observe the full gamut of time scales starting with the dynamical generation of spin-domains from an initially non-equilibrium system followed by their subsequent relaxation to progressively larger domains, i.e., coarsening.  In this model, the coalescence of domains minimizes the attractive energy by maximizing the spatial extent of regions with the same spin.  Indeed, the ground state consists of just two domains -- one for each spin  -- thereby reducing to one the number of domain walls.  The domains increase in size very slowly in time, but due to the overall decrease in the BEC's atom number, we cannot distinguish between different functional forms.

\begin{acknowledgements}
We appreciate discussions with and terminology introduced by N.~Bray-Ali, and insight gleaned from C.~Raman.  We acknowledge the financial support the NSF through the Physics Frontier Center at JQI, and the ARO with funds from both the Atomtronics MURI and DARPA's OLE Program.
\end{acknowledgements}

\bibliography{QuenchDomains}
\end{document}